\documentclass[journal]{IEEEtran}
\IEEEoverridecommandlockouts 
\usepackage[dvips]{graphicx}
\usepackage{color}
\usepackage{times}
\usepackage{fancyhdr}
\usepackage{amssymb,amsmath}
\usepackage{lastpage}
\usepackage{epsfig}
\usepackage[lined,boxed,commentsnumbered, ruled]{algorithm2e}
\usepackage[compress]{cite}
\usepackage{cases}
\usepackage{subfigure,enumitem}
\usepackage{caption}
\usepackage{longtable}
\usepackage{multirow}
\usepackage{threeparttable}
\usepackage{rotating}
\usepackage{bm,soul}
\usepackage{float}
\usepackage{tikz}
\usepackage{array}
\usepackage{subfig}

\newtheorem{theorem}{Theorem}
\newtheorem{lemma}{Lemma}
\newtheorem{corollary}{Corollary}
\newtheorem{defn}{Definition}

\newcolumntype{C}[1]{>{\centering\let\newline\\\arraybackslash\hspace{0pt}}m{#1}}

\allowdisplaybreaks[4]

\begin{document}

\title{On the Zero-Error Capacity of Semantic Channels with Input and Output Memories}
\author{Qi Cao, Yulin Shao, Shangwei Ge
	\thanks{Q. Cao  is with Xidian-Guangzhou Research Institute, Xidian University, Guangzhou, China (email: caoqi@xidian.edu.cn).}
	\thanks{Y. Shao and S. Ge are with the State Key Laboratory of Internet of Things for Smart City and the Department of Electrical and Computer Engineering, University of Macau, Macau S.A.R. Y. Shao is also with the Department of Electrical and Electronic Engineering, Imperial College London, London SW7 2AZ, U.K. (E-mails: \{ylshao,mc35283\}@um.edu.mo).}\vspace{-0.4cm}
	}
\maketitle



\begin{abstract}
This paper investigates the zero-error capacity of channels with memory. Motivated by the nuanced requirements of semantic communication that incorporate memory, we advance the classical enlightened dictator channel by introducing a new category known as the semantic channel. We analyze the zero-error capacity of the semantic channel using a comprehensive framework that accommodates multiple input and output memories. Our approach reveals a more sophisticated and detailed model compared to the classical memory channels, highlighting the impact of memory on achieving error-free communication. 
\end{abstract}

\section{Introduction}\label{sec:introduction}
The concept of channel capacity, pivotal in information theory, represents the upper limit of the communication rate over a channel. This foundational metric was first defined and thoroughly characterized by Claude Shannon in 1948 \cite{shannon1948}, marking a critical milestone in our understanding of communication systems. The pursuit of determining channel capacities across varied communication contexts has remained a central endeavor in the field for decades.
In 1956, Shannon introduced the notion of zero-error capacity \cite{shannon1956}, a measure of a channel's ability to transmit information with absolute certainty, without any errors. Unlike vanishing error capacities where errors decrease asymptotically, zero-error capacity requires that the probability of error be precisely zero. This stringent condition makes studying zero-error capacities notably challenging yet equally critical, especially for applications demanding ultra reliability \cite{park2022extreme,shao2021federated}.

The complexity of zero-error capacity arises from its combinatorial nature. Shannon's initial exploration utilized a graphical approach to represent channels; each vertex represented an input symbol, and edges connected indistinguishable outputs. He notably analyzed the noisy typewriter channel using a cycle graph with five vertices, establishing only a lower bound for its zero-error capacity. The full solution, including upper bounds, was later extended by Lov\'{a}sz in 1979 \cite{lovasz1979}. Yet, many aspects of zero-error capacity, such as for channels represented by cycle graphs of seven or more vertices, remain unresolved.

The study of zero-error capacities in channels with memory was first systematically introduced in \cite{Cai1998}. This pioneering work expanded the foundational zero-error capacity problems identified by Shannon by incorporating memory elements, thereby creating a new category of zero-error memory capacity problems. This approach not only generalized the existing theory but also facilitated the precise calculation of zero-error capacities in scenarios where only a single pair of channel elements are distinguishable. Building upon this groundwork, \cite{Cohen2016} conducted a detailed analysis of zero-error capacities for channels with a single input memory, focusing on configurations that allowed for three pairs of vertex-distinguishable elements.
Continuing this line of inquiry, subsequent research \cite{qi2018} resolved all remaining zero-error capacity challenges for binary channels with one input memory. Furthermore, in a recent advancement, \cite{qi2023} introduced the concept of the ``Chemical Residual Channel'', characterized by a single {\it output} memory that results from chemical residue effects. This innovative channel model considers how the past outputs affect the current output, and detailed computations of its zero-error capacity have shed new light on the dynamic interplay of input and memory in communication systems.

Previous research \cite{Cohen2016,qi2018} demonstrated that channels with a single memory can still be effectively analyzed using confusability graphs. This approach allows for a clear representation and computation of zero-error capacities in simpler memory configurations. However, as we expand to channels with more extended memory sequences, traditional graphical or list-based representations become inadequate due to the increased complexity and combinatorial possibilities.

An advancement in the study of long memory channels was achieved in \cite{Cai1998} with the introduction of the unique ``Enlightened Dictator Channel''. This distinctive channel type incorporates $K-1$ input memories, where the output generally mirrors the input sequence. However, when the input sequence contains $K-1$ consecutive identical inputs followed by a different input, the channel introduces a unique mechanism to handle this anomaly: it attributes a $50\%$ probability of error to this abrupt change, reflecting uncertainty about the new input. Their calculations highlight the challenges and intricacies involved in calculating zero-error capacities for channels endowed with substantial memory lengths.

{\it Contributions:} In this paper, we generalize the concept of the enlightened dictator channel and introduce a novel category of channels with memory, dubbed the ``semantic channel''. This new channel type is inspired by the unique attributes of semantic communication that incorporate memory functions \cite{sem1,sem2}. Represented by $M_{K_1,K_2}$, the semantic channel features $K_1$ input memories and $K_2$ output memories.
Specifically, when $K_2=1$, the semantic channel degenerates to the enlightened dictatorial channel;
when $K_1=1$, it degenerates into a chemical residual channel with longer memories;
when $K_1>1, K_2>1$, introduce channels with extensive memory capabilities, where the transition probabilities are influenced by both the historical inputs and outputs. This comprehensive model allows us to rigorously analyze and characterize the zero-error capacity of the semantic channel, marking a significant step forward in the theoretical understanding of memory-dependent communication systems.

{\it Notations:} Given a binary sequence $\bm{s}=s_1s_2...s_{n}$, for any $n_1,n_2\in\mathbb{Z}$ and $1\le n_1\le n_2\le n$, we define 
$\bm{s}_{n_1:n_2}\triangleq s_{n_1}s_{n_1+1}...s_{n_2}$,
$\bm{s}_{n_1}\triangleq s_{1}s_{2}...s_{n_1}$.
Let $x^{\ell}$ denote a sequence of
$x$'s of length ${\ell}$, where $x\in\{0,1\}$ and ${\ell}\ge 1$.




\section{Problem Formulation}\label{sec:problem-formulation}
\subsection{Channel with memory}
\begin{figure}[t]
  \centering
  \includegraphics[width=1\columnwidth]{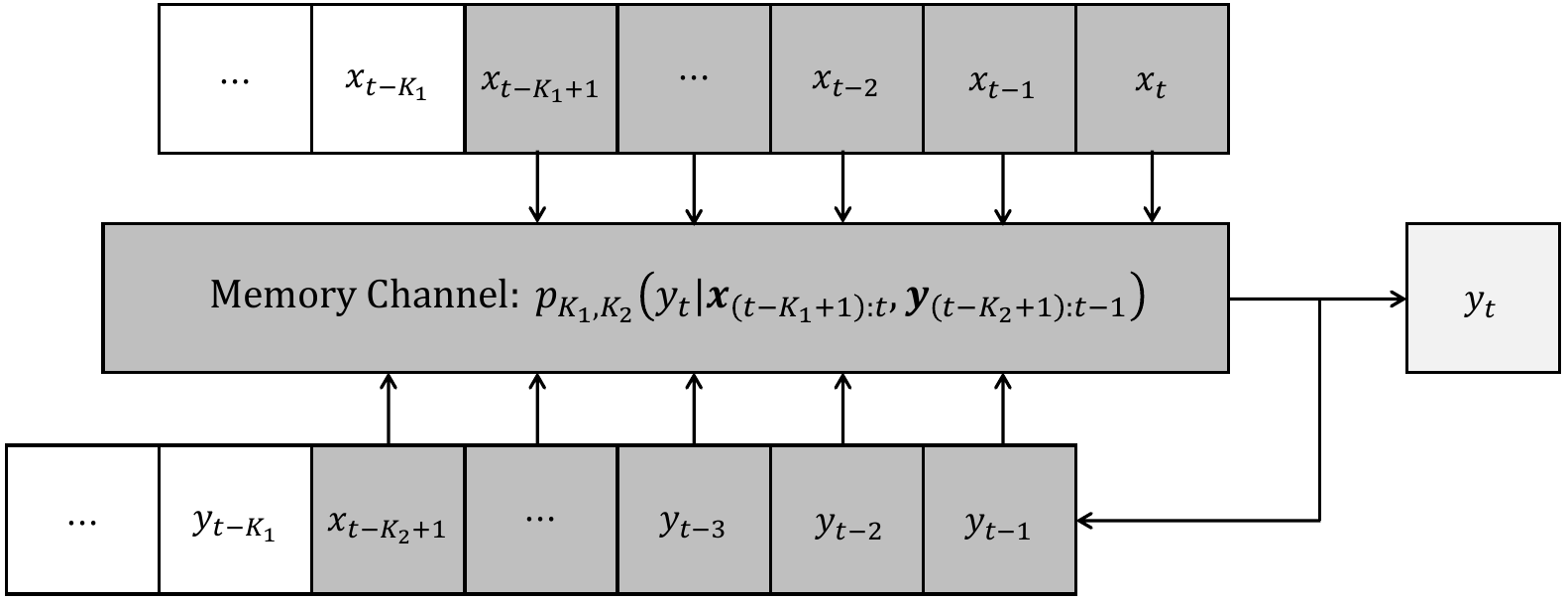}
  \caption{The output $y_t$ of a memory channel is probabilistically determined by the current input, the previous $K_1-1$ inputs, and the previous $K_2-1$ outputs.}
\label{fig:model}
\end{figure}

A channel with memory, or memory channel, possesses both input and output memory (e.g., infinite impulse response channels). In such a channel, the output is probabilistically determined by the current input, the previous $K_1-1$ inputs, and the previous $K_2-1$ outputs. As depicted in Fig.~\ref{fig:model}, at any epoch $t$, the channel input and output are denoted by $x_t$ and $y_t$, respectively, where $x_t,y_t\in\mathcal{X}\triangleq\{0,1\}$. The transition probabilities of the memory channel are represented by $p_{K_1,K_2}$. For a sufficiently large $t$ such that $t>K_1$ and $t>K_2$, we have
\begin{equation*}
    p_{K_1, K_2}\big(y_t|{\bm{x}},{\bm{y}}\big)=\Pr\big(y_t|{\bm{x}}_{(t-K_1+1):t},{\bm{y}}_{(t-K_2+1):(t-1)}\big),
\end{equation*}
where $\bm{x}\in \mathcal{X}^t$ and $\bm{y}\in \mathcal{X}^t$ are the input and output sequences of the memory channel.
In particular,
\begin{itemize}
    \item When $K_1=K_2=1$, we have $p_{1,1}(y_t|{\bm{x}},{\bm{y}})=p_{1,1}(y_t|x_t)$. The memory channel degenerates into a conventional discrete memoryless channel (DMC).
    \item When $K_1>1$ and $K_2=1$, we have $p(y_t|{\bm{x}},{\bm{y}})=p(y_t|{\bm{x}})$.  This configuration represents a channel with input memory only. The extent of this memory is determined by the value of $K_1$. 
    \item When $K_1=1$ and $K_2>1$, we have $p(y_t|{\bm{x}},{\bm{y}})=p(y_t|{\bm{y}})$. This configuration represents a channel with output memory only. The extent of this memory is determined by the value of $K_2$. 
    \item When $K_1>1$ and $K_2>1$, we have $p(y_t|{\bm{x}},{\bm{y}})=\Pr\big(y_t|{\bm{x}}_{(t-K_1+1):t},{\bm{y}}_{(t-K_2+1):(t-1)}\big)$. This configuration represents a channel with both input memory and output memory. The extent of these memories depends on the values of $K_1$ and $K_2$. 
\end{itemize}

Unlike DMCs, the zero-error capacity of memory channels requires a more intricate analysis as the confusability of symbols can depend on sequences of inputs and outputs rather than individual symbols.
Previous research has been limited to channels with a single memory, while the general problem of zero-error capacity for channels with multiple memories remains open. In this paper, we make an advancement by studying a unique class of channels with multiple memories, which we call semantic channels.

The semantic channel, denoted by $M_{K_1,K_2}$, has transition probabilities defined as
\begin{equation}\label{fml1}
    \begin{aligned}
        &\ p_{K_1,K_2}(y_t|{\bm{x}},{\bm{y}})\\       
       =&\begin{cases}
		\frac{1}{2}, & \text{if (a) $x_t\neq x_{t-1}=x_{t-2}=...=x_{t-K_1+1}$,  $t\geq K_1$},\\
            \frac{1}{2}, & \text{if (b) $x_t\neq y_{t-1}=y_{t-2}=...=y_{t-K_2+1}$, $t\geq K_2$},\\
            1, & \text{if $y_t=x_t$, and not (a), not (b)}.
		 \end{cases}        
    \end{aligned}
\end{equation}

The semantic channel is inspired by the concept of semantic communication and is a general case of the enlightened dictator channel \cite{Cai1998}. Intuitively, the semantic channel is error-free except for the two conditions stated in the first two lines of $p_{K_1,K_2}(y_t|{\bm{x}},{\bm{y}})$. To be more specific, the channel output $y_t$ will be randomly chosen from $0$ and $1$ if:
\begin{enumerate}
    \item The channel inputs have been the same (e.g., 0) for the last $K_1-1$ successive epochs, but a different input (e.g., $x_t=1$) is observed in the current slot.
    \item The channel outputs have been the same (e.g., 0) for the last $K_2-1$ successive epochs, but a different input (e.g., $x_t=1$) is observed in the current slot.
\end{enumerate}

To illustrate the motivation behind this channel, consider an analogy from semantic communication \cite{sem1,sem2} inspired by applications like event detection in sensor networks\cite{event} and anomaly detection in security systems\cite{Anoma}. Imagine two parties, Alice and Bob: Alice provides a series of reports (inputs), and Bob interprets each one (outputs). Alice might report stable environmental readings or normal system status, and Bob interprets these as a stable state. Suddenly, Alice reports a significant deviation, such as a temperature spike or unusual movement. Bob must decide if this is a meaningful anomaly or a minor fluctuation. To make this decision, he considers both Alice’s recent reports and his own prior interpretations. The semantic channel's structure captures how past inputs (Alice’s reports) and past outputs (Bob’s interpretations) influence the current transmission. Zero-error transmission via this channel helps ensure that Bob’s interpretations remain accurate, providing a reliable response.

Two special cases of the semantic channel arise when either $K_1=1$ or $K_2=1$.
\begin{itemize}[leftmargin=0.4cm]
\item When $K_2=1$, the channel $M_{K_1,1}(K_1\geq1)$ is known as an enlightened dictator channel \cite{Cai1998}. The transition probabilities for this channel are defined as
\begin{equation}
\hspace{-0.3cm} p_{K_1,1}(y_t|x_t,{\bm{x}}_{K_1\!-\!1})=\begin{cases}
			\frac{1}{2}, &\!\!\!\!\! \text{if (a) $x_t\neq {\bm{x}}_{K_1-1}$, and $t\geq K_1$},\\
            1, &\!\!\!\!\! \text{if $y_t=x_t$, and not (a)},
		 \end{cases}
\end{equation}
where $\bm{x}_{K_1-1}$ denotes the sequence $\bm{x}_{(t-K_1+1):(t-1)}$, which represents the first $K_1-1$ inputs preceding the current input $x_t$.
\item When $K_1=1$, the channel $M_{1,K_2}(K_2\geq1)$ has transition possibilities defines as
\begin{equation}\label{fml2}
\hspace{-0.3cm} p_{1,K_2}(y_t|x_t,{\bm{y}}_{K_2\!-\!1})=\begin{cases}
			\frac{1}{2}, &\!\!\!\!\! \text{if (b) $x_t\neq {\bm{y}}_{K_2-1}$, and $t\geq K_2$},\\
            1, &\!\!\!\!\! \text{if $y_t$=$x_t$, and not (b)},
		 \end{cases}
\end{equation}
where $\bm{y}_{K_2-1}$ denotes the sequence $\bm{y}_{(t-K_2+1):(t-1)}$, which represents the first $K_2-1$ outputs preceding the current input $x_t$.
\end{itemize}


\subsection{Zero-error capacity}
The concept of zero-error capacity revolves around finding the maximum communication rate at which information can be transmitted over a channel with zero probability of error. This involves constructing zero-error codes, which are sets of input sequences that can be uniquely distinguished at the output with zero error probability.
Consider a semantic channel $M_{K_1,K_2}$, where $K_1\geq 1$ and $K_2\geq 1$, with an input sequence $\bm{x}$, an output sequence $\bm{y}$, and a current input symbol $x_t\in \mathcal{X}$. We define the set of possible outputs at epoch $t$ as
\begin{equation*}
    \mathcal{O}_{K_1,K_2}(x_t,{\bm{x}},{\bm{y}})\triangleq\{y_t: p_{K_1,K_2}(y_t|x_t,{\bm{x}},{\bm{y}})>0\}.
\end{equation*}
Any $y\in \mathcal{O}_{K_1,K_2}(x,{\bm{x}}_{K_1-1},{\bm{y}}_{K_2})\subseteq  \{0,1\}$ is called a \emph{possible output} when the current input is $x$ and the  previous input and output sequences are $\bm{x}$ and $\bm{y}$, respectively.

When the input is sequence $\bm{x}=(x_1,x_2,\dots,x_n)\in \mathcal{X}^n$, we further define a set $\mathcal{O}_{K_1,K_2}(\bm{x})$ to represent the set of all possible output sequences:
\begin{equation*}
\mathcal{O}_{K_1,K_2}(\bm{x})\triangleq\{\bm{y}\in\mathcal{X}^n \!:\!y_t\!\in\!\mathcal{O}_{K_1,K_2}(x_t,{\bm{x}}_{1:{K_1\!-\!1}},{\bm{y}}_{1:{K_2\!-\!1}})\}.
\end{equation*}

\begin{defn}[distinguishable sequences]\label{df1}
The sequences $\bm{x}, \bm{x}'\in\mathcal{X}^n$ are \emph{distinguishable} for the channel~$M_{K_1,K_2}$ if 
	\begin{equation}
	\mathcal{O}_{K_1,K_2}(\bm{x})\cap \mathcal{O}_{K_1,K_2}(\bm{x}')=\emptyset.
	\end{equation}
\end{defn}

\begin{defn}[zero-error codes]\label{df2}
	Let $\mathcal{B}_n\subseteq \mathcal{X}^n$ be a set of length-$n$ sequences. $\mathcal{B}_n$ is called a \emph{code} of length $n$ for a semantic channel $M_{K_1,K_2}$ with transition possibility $p(y_t|x_t,{\bm{x}}_{K_1-1},{\bm{y}}_{K_2-1})$ if any two distinct sequences in $\mathcal{B}_n$ are distinguishable for $M_{K_1,K_2}$. That is,
	\begin{align}
	    \mathcal{O}_{K_1,K_2}(\bm{x})\cap \mathcal{O}_{K_1,K_2}(\bm{x}')=\emptyset, \;\forall\, \bm{x},\bm{x}'\in \mathcal{B}_n.
	\end{align}
	The sequences in $\mathcal{B}_n$ are called {\em codewords}.
\end{defn}

\begin{defn}[zero-error capacity]
	A zero-error code $\mathcal{B}^*_n$ of length $n$ for the channel $M_{K_1,K_2}$ is said to be \emph{optimal} if $\mathcal{B}^*_n$ achieves the largest cardinality of a code of length $n$ for the channel $M_{K_1,K_2}$.
The maximal achievable rate of a length-$n$ code is defined as
\begin{equation}
	    R_n\triangleq\frac{1}{n}\log |\mathcal{B}^*_n|.
\end{equation}
The \emph{zero-error capacity} of $M_{K_1,K_2}$ is defined as
\begin{equation}\label{eq:capacity}
    C(M_{K_1,K_2})\triangleq\limsup\limits_{n\to \infty}R_n.
\end{equation}
\end{defn}

Note that the base of the logarithm in \eqref{eq:capacity} is $2$, and will be omitted throughout this paper.
Since  $1 \le|\mathcal{B}^*_n|\le |\mathcal{X}|^{n}=2^n$, we have  $0\le C(M_{K_1,K_2})\le 1$. 

In the example with Alice and Bob, each report Alice sends represents a codeword in the channel, and all possible reports form a zero-error code. The zero-error property ensures that each codeword transmitted has a unique interpretation.

\section{Main Results}\label{sec:main-results}
In this section, we analyze the zero-error capacity of the semantic channel $M_{K_1,K_2}$. Our analysis is divided into three parts, each addressing different cases: $K_2=1$, $K_1=1$, and the $K_1,K_2\geq 2$ case. This allows us to provide a comprehensive understanding of the zero-error capacity for each specific scenario.

\begin{lemma}\label{lm}
Consider a zero-error code $\mathcal{C}_n$ of length $n$ for a semantic channel. Let $\bm{x}$ be an arbitrary but fixed codeword in $\mathcal{C}_n$. For any sequence $\bm{x}'\in \mathcal{X}^n$ such that $\mathcal{O}_{K_1,K_2}(\bm{x}')\subseteq \mathcal{O}_{K_1,K_2}(\bm{x})$, by replacing $\bm{x}$ by $\bm{x}'$, the updated $\mathcal{C}_n$ remains to be a zero-error code for the channel.
\end{lemma}
\begin{IEEEproof}
    It suffices to prove that any sequence $\bm{y}\in \mathcal{C}_n\setminus \{\bm{x}\}$ is distinguishable from $\bm{x}'$. Since $\mathcal{C}_n$ is a code, we have $\mathcal{O}_{K_1,K_2}(\bm{x})\cap\mathcal{O}_{K_1,K_2}(\bm{y})=\emptyset$. On the other hands, $\mathcal{O}_{K_1,K_2}(\bm{x}')\subseteq \mathcal{O}_{K_1,K_2}(\bm{x})$.
    Therefore, 
    \begin{equation*}
        \mathcal{O}_{K_1,K_2}(\bm{x}')\cap\mathcal{O}_{K_1,K_2}(\bm{y})\subseteq \mathcal{O}_{K_1,K_2}(\bm{x})\cap\mathcal{O}_{K_1,K_2}(\bm{y}) =\emptyset.
    \end{equation*}
    That is, $\bm{x}'$ is distinguishable from $\bm{y}$.
\end{IEEEproof}
\begin{corollary}\label{cor}
    Given two semantic channels $A$ and $B$, if any $\bm{x}\in \mathcal{X}^n$ satisfies that $\mathcal{O}_{A}(\bm{x})\subseteq \mathcal{O}_{B}(\bm{x})$, then $C(A)\ge C(B)$.
\end{corollary}

\subsection{The $K_2=1$ case}
To start with, we investigate the zero-error capacity of $M_{K_1,1}$. This channel has been previously studied under the name of the enlightened dictator channel \cite{Cai1998}. The authors obtained the following result.
\begin{theorem}[zero-error capacity of $M_{K_1,1}$ \cite{Cai1998}]\label{thm:cai}
    \begin{equation}
        C(M_{K_1,1})= \log \lambda_{K_1},
    \end{equation}
    where $\lambda_{K_1}$ is the largest real root of $\lambda^{K_1-1}-\sum_{j=0}^{K_1-2}\lambda^j=0$.
\end{theorem}

However, we find that the authors in \cite{Cai1998} omitted the $K_1=2$ case, and Theorem \ref{thm:cai} is valid only when $K_1\geq 3$. Theorem \ref{thm:1} below fills this gap. 

\begin{theorem}\label{thm:1}
The zero-error capacity of $M_{2,1}$ is
\begin{equation}
    C(M_{2,1})=\frac{1}{2}.
\end{equation}
\end{theorem}
\begin{IEEEproof}
Consider a sequence of sets $\{\mathcal{C}_n\}$ such that
\begin{align}\label{eq:C_n}
\mathcal{C}_n=\begin{cases}
	\{00,11\}^{n/2}, & \text{if $n$ is even,}\\
        \{0,1\}\times \{00,11\}^{(n-1)/2}, & \text{if $n$ is odd.}
	\end{cases}
\end{align}
The asymptotic rate of $\mathcal{C}_n$ is $1/2$.

Let $\bm{x},\bm{x}'\in \mathcal{C}_n$ be arbitrary and $\bm{y}\in \mathcal{O}_{2,1}(\bm{x})$ and $\bm{y}'\in \mathcal{O}_{2,1}(\bm{x}')$ be arbitrary.
When $n$ is even, there exists at least one coordinate $i$ such that $\{x_ix_{i+1},x'_ix'_{i+1}\}= \{00,11\}$, i.e., $x_i=x_{i+1}\neq x'_i=x'_{i+1}$.
We further have $p(y_{i+1}=x_{i+1}|x_{i+1},x_i)=p(y'_{i+1}=x'_{i+1}|x'_{i+1},x'_i)=1$, and thus $y_{i+1}\neq y'_{i+1}$. In other words, for any two sequences in $\mathcal{O}_{2,1}(\bm{x})$ and $\mathcal{O}_{2,1}(\bm{x}')$, respectively, the $i+1$-th bits are different. Hence, $\mathcal{O}_{2,1}(\bm{x})\cap\mathcal{O}_{2,1}(\bm{x}')=\emptyset$, and then $\bm{x}$ and $\bm{x}'$ are distinguishable for $M_{2,1}$. We can obtain that $\{\mathcal{C}_n\}$ is a sequence of codes for $M_{2,1}$ when $n$ is even.
Likewise, when $n$ is odd, $x_1\neq x'_1$ or there exists at least one coordinate $i$ such that $\{x_ix_{i+1},x'_ix'_{i+1}\}= \{00,11\}$.
Then we can obtain that $y_{i+1}\neq y'_{i+1}$, and thus $\{\mathcal{C}_n\}$ is a sequence of codes for $M_{2,1}$ when $n$ is odd. Therefore, 
$C(M_{2,1})\ge \frac{1}{2}.$

Now we prove the converse part. It suffices to prove that any code $\mathcal{A}_n$ for $M_{2,1}$ satisfies that $|\mathcal{A}_n|\le |\mathcal{C}_n|$.
For any $\bm{a}\in \mathcal{A}_n$, let 
\begin{equation}\label{eq:I_max}
    I_{\bm{a}} \triangleq\max\{n-2j:a_{n-2j}\neq a_{n-2j-1}\}.
\end{equation}
Let $\bm{x}\in \mathcal{A}_n$ be such that
$I_{\bm{x}} =\max\{I_{\bm{a}}:\bm{a}\in \mathcal{A}_n \}$.
Let $\bm{x}'$ be a sequence of length $n$ such that 
\begin{equation}\label{eq:x'_I=}
    x'_{I_{\bm{x}}}=x'_{I_{\bm{x}}-1}
\end{equation} and $x'_i=x_i$ for $i=\mathbb{Z}[1,n]\setminus \{I_{\bm{x}}\}$. 
Now we prove that by replacing $\bm{x}$ in $\mathcal{A}_n$ by $\bm{x}'$, the updated $\mathcal{A}_n$ remains code.
For ${I_{\bm{x}}}=n$, we can see $\mathcal{O}_{2,1}(\bm{x}')\subseteq \mathcal{O}_{2,1}(\bm{x})$. Thus, the updated $\mathcal{A}_n$ remains a code.

Now we prove that for ${I_{\bm{x}}}\in\mathbb{Z}[2,n-2]$, the updated $\mathcal{A}_n$ also remains a code. Assume the contrary, there exists a codeword $\bm{z}\in \mathcal{A}_n\setminus\{x\}$ such that $\bm{z}$ is distinguishable from $\bm{x}$ but not $\bm{x}'$.
Since $x_{i-1}x_i=x'_{i-1}x'_i$ for any $i\in \mathbb{Z}[1,n]\setminus \{I_{\bm{x}}, I_{\bm{x}}+1\}$, there exists a sequence $\bm{y}\in \mathcal{O}_{2,1}(\bm{x})$ such that $y'_i=y_i$ for $i\in \mathbb{Z}[1,n]\setminus \{I_{\bm{x}}, I_{\bm{x}}+1\}$.
Further, by~\eqref{eq:x'_I=}, we can see $y'_{I_{\bm{x}}}=x'_{I_{\bm{x}}}$. While $x_{I_{\bm{x}}}\neq x_{I_{\bm{x}}-1}$ implies that $y_{I_{\bm{x}}}$ could be equal to $0$ or $1$. 
Therefore, $\bm{z}$ is distinguishable from $\bm{x}$ but not $\bm{x}'$ at coordinate $I_{\bm{x}}+1$, and thus
$z_{I_{\bm{x}}}=z_{I_{\bm{x}}+1}\neq x_{I_{\bm{x}}+1}=x_{I_{\bm{x}}}=x'_{I_{\bm{x}}+1}$.
By~\eqref{eq:I_max} and \eqref{eq:x'_I=}, we further have
$z_{I_{\bm{x}}+2}=z_{I_{\bm{x}}+1}\neq x'_{I_{\bm{x}}+1}=x_{I_{\bm{x}}+1}=x_{I_{\bm{x}}+2}$.
Thus, $\mathcal{O}(\bm{z}_{I_{\bm{x}}+2})\neq \mathcal{O}(\bm{x}'_{I_{\bm{x}}+2})$, i.e., $\bm{z}$ and $\bm{x}'$ are distinguishable, which contradicts the assumption.
Therefore, the updated $\mathcal{A}_n$ always remains a code.
Thus the replacement can be repeated until for any codeword $\bm{x}$ in the final updated $\mathcal{A}_n$, 
\begin{align*}
\begin{cases}
	x_1=x_2,x_3=x_4,...,x_{n-1}=x_n, & \text{if $n$ is even,}\\
        x_2=x_3,x_4=x_5,...,x_{n-1}=x_n, & \text{if $n$ is odd.}
	\end{cases}
\end{align*}
Therefore, the final updated $\mathcal{A}_n$ is a subset of $\mathcal{C}_n$, and so $|\mathcal{A}_n|\le |\mathcal{C}_n|$.
\end{IEEEproof}

\subsection{The $K_1=1$ case}
Next, we examine the channel $C(M_{1,K_2})$ and derive its zero-error capacity.

\begin{theorem}[zero-error capacity of $M_{1,K_2}$]\label{thm:2}
\begin{equation}
C(M_{1,K_2})=
\begin{cases}
0,    & \text{if}~K_2=2, \\
 \log \omega_{K_2},    & \text{if}~K_2\geq 3.
\end{cases}
\end{equation}
where $\omega_{K_2}$ is the only positive root of the equation
\begin{equation}
    \omega^{K_2-2}-\sum_{j=0}^{K_2-3}\omega^j=0.
\end{equation}
\end{theorem}

\begin{IEEEproof}
We first consider the $K_2=2$ case.
Let $\bm{x}\in \mathcal{X}^n$ be an arbitrary input sequence. Let $\bm{y}$ be a corresponding output sequence.
First, $y_1=x_1$. We can also see that when $y_{t}=x_1$ for some positive number $t$, if $x_{t+1}=y_{t}=x_1$, then $y_{t+1}=x_{t+1}=x_1$. Otherwise, if $x_{t+1}\neq y_{t}$, then $p(y_{t+1}=x_1)=1/2$, i.e., $y_{t+1}=x_1$ is also a possible output. Thus, the sequence ${x_1}^n$ is a possible output.
Therefore, for any $\bm{x},\bm{x}'\in \mathcal{X}^n$ with $x_1=x'_1$,
$\bm{x}$ and $\bm{x}'$ are not distinguishable. Thus, $C(M_{1,2})=0$.

Next, we consider the $K_2\geq 3$ case.
Let $\mathcal{C}_n$ be a set of all sequences such that any sequence in the set does not contain the substring $0^{K_2-1}$ or $1^{K_2-1}$. 
For any $\bm{x}\in \mathcal{C}_n$,  $\bm{y}=\bm{x}$ is the only possible output. Thus $\mathcal{C}_n$ is a code for $M_{1,K_2}$.

Now we define a subset $\mathcal{C}_{n,i}$ of $\mathcal{C}_n$, for any $i\in\mathbb{Z}[1,n].$
\begin{align*}
    \mathcal{C}_{n,i}\triangleq \{\bm{x}\in\mathcal{C}_n:x_{n-i}\neq x_{n-i+1}=x_{n-i+2}=...=x_{n}\}
\end{align*}
Since any sequence in $\mathcal{C}_n$ does not contain the substring $0^{K_2-1}$ or $1^{K_2-1}$, we have
 $\mathcal{C}_{n,i}=\emptyset$ for any $i\in\mathbb{Z}[K_2-1,n]$, i.e.,
$\mathcal{C}_n=\mathcal{C}_{n,1}\cup\mathcal{C}_{n,2}\cup...\cup\mathcal{C}_{n,K_2-2}$.
For $i_1,i_2\in\mathbb{Z}[K_2-1,n]$, $i_1\neq i_2$, we have
$\mathcal{C}_{n,i_1}\cap \mathcal{C}_{n,i_2}=\emptyset.$
Thus, 
\begin{equation}\label{eq:Dnsum}
    |\mathcal{C}_{n}|=|\mathcal{C}_{n,1}|+|\mathcal{C}_{n,2}|+...+|\mathcal{C}_{n,K_2-2}|.
\end{equation}
By adding a binary suffix $s$ to each sequence $\bm{x}\in \mathcal{C}_n$, such that $s\neq x_n$, we can obtain a sequence in $\mathcal{C}_{n+1,1}$. Vice versa, i.e., by removing the last bit in each sequence in $\mathcal{C}_{n+1,1}$, a sequence in $\mathcal{C}_n$ is obtained. Therefore, 
\begin{equation}\label{eq:Dn}
    |\mathcal{C}_{n}|=|\mathcal{C}_{n+1,1}|.
\end{equation}
Likewise, by adding a binary suffix $s$ to each sequence $\bm{x}\in \mathcal{C}_{n,i}$, $i=1,2,...,K_2-3$, such that $s= x_n$, we can obtain a sequence in $\mathcal{C}_{n+1,i+1}$ and vice versa.
Thus,
\begin{equation}\label{eq:Dni}
    |\mathcal{C}_{n,i}|=|\mathcal{C}_{n+1,i+1}|\text{, for } i=1,2,...,K_2-3.
\end{equation}
By \eqref{eq:Dn} and \eqref{eq:Dni}, for $i=1,2,...,K_2-2$, we have
\begin{equation}\label{eq:Dneq}
|\mathcal{C}_{n,i}|=|\mathcal{C}_{n-1,i-1}|=...=|\mathcal{C}_{n-i+1,1}|=|\mathcal{C}_{n-i}|.
\end{equation}
By \eqref{eq:Dnsum} and \eqref{eq:Dneq}, we have $|\mathcal{C}_{n}|=\sum_{i=1}^{K_2-2}|\mathcal{C}_{n-i}|$, which implies that the asymptotic rate of $\mathcal{C}_{n}$ is 
 $\log \omega_{K_2}$.

We next prove the converse part, i.e., $C(M_{1,K_2})\leq \log \omega_{K_2}$.
Let $\{\mathcal{D}_n\}$ be a sequence of optimal codes for the channel $M_{1,K_2}$. For any codeword $\bm{x}$ in $\mathcal{D}_n$, where $n\ge k-1$, let $I_{\bm{x}}$ be a function of $\bm{x}$ such that if $\bm{x}$ does not contain $0^{K_2-1}$ or $1^{K_2-1}$, then let $I_{\bm{x}}=n$; otherwise, let 
$I_{\bm{x}}=\min\{i:x_{i-K_2+2}=x_{i-K_2+3}=...=x_{i}\}$.

If $\bm{x}$ contains the substring $s^{K_2-1}$ for some $s\in \{0,1\}$, let $i\in\mathbb{Z}[K_2-1,n-k+2]$ be the smallest coordinate such that 
\begin{equation}\label{eq:xeq}
    x_{i-K_2+2}x_{i-K_2+3}...x_{i} = s^{K_2-1}.
\end{equation} 
Let $\bm{y}$ be a sequence of length $n$ such that 
\begin{align}
y_t=\left\{
\begin{array}{ll}
	x_t, & \hbox{for $t\in \mathbb{Z}[1,i-1]$}, \\
	s, & \hbox{for  $t\in \mathbb{Z}[i,n]$.}
\end{array}
\right.
\end{align}
Now we show that $\bm{y}\in \mathcal{O}(\bm{x})$.
First, for any $t<i$, (a) in (\ref{fml2}) is not satisfied, and thus $y_t=x_t$.
For $t=i$, $x_t\neq y_{t-1}=y_{t-1}=...=y_{t-K_2+2}$, i.e., (b) in (\ref{fml2}) is satisfied. Thus, $y_t=s$ is a possible output. We can further obtain that for $t>i$, if $y_{i}=y_{i+1}=...=y_{t-1}=s$, then $y_t=s$ is a possible output.
Therefore, $\bm{y}\in \mathcal{O}(\bm{x})$. 

Now we prove that by replacing $\bm{x}$ by $\bm{y}$ in $\mathcal{D}_n$, the updated $\mathcal{D}_n$ remains a code for the channel $M_{1,K_2}$.
By Lemma~\ref{lm}, it is sufficient to prove that $\mathcal{O}_{1,K_2}(\bm{y})\subseteq \mathcal{O}_{1,K_2}(\bm{x})$. 
We can see that for any $t\in \mathbb[Z][1,n]$, (b) in (\ref{fml2}) is not satisfied. Thus, $\mathcal{O}_{1,K_2}(\bm{y})=\{\bm{y}\}\subseteq \mathcal{O}_{1,K_2}(\bm{x})$.
Therefore, the replacement can be repeated until for any codeword $\bm{x}$ in the final updated code, we have
$x_i=x_{I_{\bm{x}}}$ for any $i\ge I_{\bm{x}}$. Let $\mathcal{E}_n$ be a set of sequences not containing $0^{K_2-1}1$ or $1^{K_2-1}0$. Clearly, $|\mathcal{D}_n|\le |\mathcal{E}_n|$. 

It suffices to prove that the asymptotic rate of $\mathcal{E}_n$ is $ \log \omega_{K_2}$.
We define a subset $\mathcal{E}_{n,i}$ of $\mathcal{E}_n$, for any $i\in\mathbb{Z}[1,n].$
\begin{align*}
    \mathcal{E}_{n,i}\triangleq \{\bm{x}\in\mathcal{E}_n:x_{n-i}\neq x_{n-i+1}=x_{n-i+2}=...=x_{n}\}
\end{align*}
Clearly, 
\begin{equation}\label{eq:Cnsum}            |\mathcal{E}_{n}|=|\mathcal{E}_{n,1}|+|\mathcal{E}_{n,2}|+...+|\mathcal{E}_{n,n}|.
\end{equation}
By adding a binary suffix $s$ to each sequence $\bm{x}\in \bigcup_{i=1}^{K_2-2}\mathcal{E}_{n,i}$, such that $s\neq x_n$, we can obtain a sequence in $\mathcal{E}_{n+1,1}$. Vice versa, i.e., by removing the last bit in each sequence in $\mathcal{E}_{n+1,1}$, a sequence in $\bigcup_{i=1}^{K_2-2}\mathcal{E}_{n,i}$ is obtained. Thus,
\begin{equation}\label{eq:Cn}
    |\mathcal{E}_{n+1,1}|=\sum_{i=1}^{K_2-2}|\mathcal{E}_{n,i}|.
\end{equation}
Likewise, by adding a binary suffix $s$ to each sequence $\bm{x}\in \mathcal{E}_{n,i}$, $i=1,2,...,n$, such that $s= x_n$, we can obtain a sequence in $\mathcal{E}_{n+1,i+1}$ and vice versa.
Thus,
\begin{equation}\label{eq:Cni}
    |\mathcal{E}_{n,i}|=|\mathcal{E}_{n+1,i+1}|\text{, for } i=1,2,...,n.
\end{equation}
By~\eqref{eq:Cn} and \eqref{eq:Cni}, we have
$|\mathcal{E}_{n,1}|=\sum_{i=1}^{K_2-2}|\mathcal{E}_{n-i,1}|$.
Thus, the asymptotic rate of $|\mathcal{E}_{n,1}|$ is $\log \omega_{K_2}$.
By~\eqref{eq:Cnsum}, we have
$|\mathcal{E}_{n,1}|/n\le |\mathcal{E}_{n,1}|\le |\mathcal{E}_{n,1}|$. Therefore, the asymptotic rate of $|\mathcal{E}_{n}|$ is also $\log \omega_{K_2}$.
\end{IEEEproof}

\subsection{The $K_1,K_2\geq2$ case}
Finally, we consider the general semantic channel $M_{K_1,K_2}$ with $K_1,K_2\geq2$. This case can be further partitioned into three subcases: $K_1\ge K_2\ge 2$, $K_2>K_1> 2$, and $K_2>K_1= 2$.
For the subcase $K_1\ge K_2\ge 2$, we can determine the exact zero-error capacity. For the other two subcases, we provide bounds for the zero-error capacity.

\begin{theorem}[zero-error capacity of $M_{K_1,K_2}$, $K_1,K_2\geq2$]\label{thm:3}
\begin{itemize}
   \item If $K_2\in \{2,3\}$ or $K_1\geq K_2\geq4$,
    \begin{equation}\label{k1k22}
        C(M_{K_1,K_2})=\log\omega_{K_2}.
    \end{equation}
    \item If $K_1=2$ and $K_2>3$,
    \begin{equation}\label{k2k1}
        0\le C(M_{K_1,K_2})\le 1/2.
    \end{equation}
     \item If $K_2>K_1\ge 3$,
    \begin{equation}\label{k2k1_new}
        \log \omega_{K_1} \le C(M_{K_1,K_2}) \le \log \lambda_{K_1}.
    \end{equation}
    
\end{itemize}
\end{theorem}

\begin{figure}[t]
\centering
\begin{tikzpicture}[y=2.5cm,x=0.4cm]
\draw[xstep=1, ystep=0.5, gray!20] (0,0) grid (15,1);
\draw[->,very thick] (0,0) -- (16,0) node[right] {$K_2$};
\draw[->,very thick] (0,0) -- (0,1.2) node[left] {Capacity};
\foreach \i in {1,2,...,15} {\draw (\i,0) -- +(0,2pt);}
\foreach \i in {0,0.1,0.2,...,1.0} {\draw (0,\i)  -- +(2pt,0);}
\foreach \i in {3,6,...,15} {\draw (\i,0) node[below]{\i} -- +(0,2pt);}
\foreach \i in {0,0.5,1} {\draw (0,\i) node[left]{\i} -- +(2pt,0);}

\foreach \x/\y in {2/0, 3/0, 4/0.6942, 5/0.8791, 6/0.9468, 7/0.9752, 8/0.9881, 9/0.9942, 10/0.9971, 11/0.9986, 12/0.9993, 13/0.9996, 14/0.9998, 15/0.9999} {\fill[cyan] (\x,\y) circle (2pt);}
\draw[thick,cyan] plot coordinates {(2,0) (3,0) (4,0.6942) (5,0.8791) (6,0.9468) (7,0.9752) (8,0.9881) (9,0.9942) (10,0.9971) (11,0.9986) 
    (12,0.9993) (13,0.9996) (14,0.9998) (15,0.9999)};
\end{tikzpicture}
\caption{Zero-error capacity of the semantic channel when $K_1=1$ or $K_2= 2,3$ or $K_1\ge K_2\ge 4$.}\label{fig:solved}
\end{figure}

\begin{figure}[t]
\centering
\begin{tikzpicture}[y=2.5cm,x=0.4cm]
\draw[xstep=1, ystep=0.5, gray!20] (0,0) grid (15,1);
\draw[->,very thick] (0,0) -- (16,0) node[right] {$K_1$};
\draw[->,very thick] (0,0) -- (0,1.2) node[left] {Capacity};
\foreach \i in {1,2,...,15} {\draw (\i,0) -- +(0,2pt);}
\foreach \i in {0,0.1,0.2,...,1.0} {\draw (0,\i)  -- +(2pt,0);}
\foreach \i in {3,6,...,15} {\draw (\i,0) node[below]{\i} -- +(0,2pt);}
\foreach \i in {0,0.5,1} {\draw (0,\i) node[left]{\i} -- +(2pt,0);}

\foreach \x/\y in {2/0.5000, 3/0.6942, 4/0.8791, 5/0.9468, 6/0.9752, 7/0.9881, 8/0.9942, 9/0.9971, 10/0.9986, 11/0.9993, 12/0.9996, 13/0.9998, 14/0.9999, 15/1.0000} {\fill[blue] (\x,\y) circle (2pt);}
\draw[thick,blue] plot coordinates {(2,0.5000) (3,0.6942) (4,0.8791) (5,0.9468) (6,0.9752) (7,0.9881) (8,0.9942) (9,0.9971) (10,0.9986) 
    (11,0.9993) (12,0.9996) (13,0.9998) (14,0.9999) (15,1.0000)};

\foreach \x/\y in {2/0, 3/0, 4/0.6942, 5/0.8791, 6/0.9468, 7/0.9752, 8/0.9881, 9/0.9942, 10/0.9971, 11/0.9986, 12/0.9993, 13/0.9996, 14/0.9998, 15/0.9999} {\fill[red] (\x,\y) circle (2pt);}
\draw[thick,red] plot coordinates {(2,0) (3,0) (4,0.6942) (5,0.8791) (6,0.9468) (7,0.9752) (8,0.9881) (9,0.9942) (10,0.9971) (11,0.9986) 
    (12,0.9993) (13,0.9996) (14,0.9998) (15,0.9999)};
    
\draw[thick,blue] (10,0.4) -- (11,0.4) node[right,black] {\footnotesize upper bound};
\draw[thick,red] (10,0.2) -- (11,0.2) node[right,black] {\footnotesize lower bound};
\fill[blue] (10.5,0.4) circle (2pt);
\fill[red] (10.5,0.2) circle (2pt);

\end{tikzpicture}
\caption{An illustration of the upper and lower bounds in \eqref{k2k1_new}.}
\label{fig:Bounds}
\end{figure}

\begin{figure}[t]
\centering
\begin{tikzpicture}[y=0.3cm,x=0.3cm]
\draw[step=1, gray!20] (0,0) grid (15,15);
\draw[->,very thick] (0,0) -- (16,0) node[right] {$K_1$};
\draw[->,very thick] (0,0) -- (0,16) node[left, above] {$K_2$};
\foreach \i in {1,2,...,15} {\draw (\i,0) -- +(0,2pt);}
\foreach \i in {1,2,...,15} {\draw (0,\i)  -- +(2pt,0);}
\foreach \i in {3,6,...,15} {\draw (\i,0) node[below]{\i} -- +(0,2pt);}
\foreach \i in {3,6,...,15} {\draw (0,\i) node[left]{\i} -- +(2pt,0);}

\foreach \x in {1,2,...,15} {
    \foreach \y in {1,2,...,15} {
        \ifnum\x<2
            \node[draw,circle,inner sep=1.5pt] at (\x,\y) {}; 
        \else
            \ifnum\y<\x
                \node[draw,circle,inner sep=1.5pt] at (\x,\y) {}; 
            \else
                \ifnum\y<4
                    \node[draw,circle,inner sep=1.5pt] at (\x,\y) {}; 
                \else
                    \ifnum\y=\x
                        \node[draw,circle,inner sep=1.5pt] at (\x,\y) {}; 
                    \else
                        \fill[black] (\x,\y) circle (2pt); 
                    \fi
                \fi
            \fi
        \fi
    }
}
\end{tikzpicture}
\caption{The hollow points represent the cases where the zero-error capacities are determined. The solid points represent the cases where the zero-error capacities are bounded.}
\label{fig:all}
\end{figure}

\begin{IEEEproof}
We first derive a general upper bound on $C(M_{K_1,K_2})$.
Note that any $\bm{x}\in \mathcal{X}^n$ satisfies $$\mathcal{O}_{1,K_2}(\bm{x})\subseteq \mathcal{O}_{K_1,K_2}(\bm{x})\text{ and }\mathcal{O}_{K_1,2}(\bm{x})\subseteq \mathcal{O}_{K_1,K_2}(\bm{x}).$$ 
By Corollary~\ref{cor}, we have
\begin{equation}\label{equpperbound}
    C(M_{K_1,K_2})\le \min\{C(M_{K_1,1}),C(M_{1,K_2})\}.
\end{equation}

Now we prove that \eqref{k1k22} is true. When $K_2$ equals $2$ or $3$, we have $0\le C(M_{K_1,K_2})\le \min\{C(M_{K_1,1}),C(M_{1,K_2})\}=0$;
and hence, $C(M_{K_1,K_2})=0=\log \omega_{K_2}$.
When $4\le K_2\le K_1$, 
let $\{\mathcal{C}_n\}$ be a sequence of sets such that that $\mathcal{C}_n$ is a set of sequences not containing substring $0^{K_2-1}$ or $1^{K_2-1}$.
For any input sequence $\bm{x}\in \mathcal{C}_n$, $\bm{x}$ itself is the only possible output in $\mathcal{O}_{K_1,K_2}(\bm{x})$. Thus $\mathcal{C}_n$ is a code for $C(M_{K_1,K_2})$. By the proof of Theorem~\ref{thm:3}, the asymptotic rate of $\{\mathcal{C}_n\}$ is $\log \omega_{K_2}$.
By \eqref{equpperbound}, we can further obtain
\begin{align*}
\log \omega_{K_2}\le &~C(M_{K_1,K_2})\\
\le &~\min\{C(M_{K_1,1}),C(M_{1,K_2})\}\\
\le &~C(M_{1,K_2})=\log \omega_{K_2}.
\end{align*} 
Therefore, \eqref{k1k22} holds when $K_1\ge K_2\ge 2$.

Next, we consider the remaining two cases. If $K_2=2$, clearly $C(M_{K_1,K_2})\ge 0$.
If $K_2>2$, let $\{\mathcal{C}_n\}$ be a sequence of sets such that that $\mathcal{C}_n$ is a set of sequences not containing substring $0^{K_1-1}$ or $1^{K_1-1}$. 
Let $\bm{x}$ be an arbitrary sequence in $\{\mathcal{C}_n\}$.
We can see that neither (a) nor (b) in \eqref{fml1} is satisfied when the input is $\bm{x}$.
Thus,
$$\mathcal{O}_{K_1,K_2}(\bm{x})=\{\bm{x}\},$$ which implies that $\mathcal{C}_n$ is a code for $M_{K_1,K_2}$.
We can further obtain the asymptotic rate of $\{\mathcal{C}_n\}$ is $\log \omega_{K_1}$, and thus $C(M_{K_1,K_2})\ge \log \omega_{K_1}$.

For the upper bound, by \eqref{equpperbound}, we have
\begin{align*}
C(M_{K_1,K_2})\le &~\min\{C(M_{K_1,1}),C(M_{1,K_2})\}\\
\le &~C(M_{K_1,1})=\log \lambda_{K_1}.
\end{align*} 
Therefore, \eqref{k2k1} holds if $K_1= 2$ and $K_2>3$, and \eqref{k2k1_new} holds if $K_2>K_1\ge 3$.
\end{IEEEproof}

Fig.~\ref{fig:solved} presents the zero-error capacity of the semantic channel for cases where $K_1=1$ or $K_2= 2,3$ or $K_1\ge K_2\ge 4$. An illustration of the bounds in \eqref{k2k1} and \eqref{k2k1_new} is given in Fig. \ref{fig:Bounds}. As can be seen, with the increase in $K_1$, the upper and lower bounds of $C(M_{K_1,K_2})$ gradually converge.
Additionally, Fig.~\ref{fig:all} provides a comprehensive illustration of both determined zero-error capacities, and undetermined but bounded zero-error capacities.

\section{Conclusion}\label{sec:conclusion}
This paper extended the theoretical framework of zero-error capacity to include channels with both (long) input and output memories. By introducing the concept of semantic channels, we advance beyond the classical enlightened dictator channel model. Through rigorous analysis, we revealed how memory influence error-free communication rate, providing richer and deeper perspective than traditional approaches.
Our future work will focus on extending our approach to accommodate a broader range of memory configurations, further enhancing the robustness and applicability of our theoretical constructs.

\bibliographystyle{IEEEtran}
\bibliography{References}

\end{document}